\documentclass[acmlarge,manuscript,screen]{acmart}

\usepackage{graphicx}
\usepackage{multirow}
\usepackage{adjustbox}
\usepackage{longtable}
\usepackage{caption}

\AtBeginDocument{%
  \providecommand\BibTeX{{%
    \normalfont B\kern-0.5em{\scshape i\kern-0.25em b}\kern-0.8em\TeX}}}

\setcopyright{acmlicensed}
\acmJournal{PACMHCI}
\acmYear{2025} 

\begin{document}

\title{A Systematic Literature Review of Infrastructure Studies in SIGCHI}

\author{Yao Lyu}
\orcid{0000-0003-3962-4868}
\affiliation{%
  \institution{University of Michigan}
  \city{Ann Arbor}
  \state{Michigan}
  \country{USA}
}
\email{yaolyu@umich.edu}

\author{Jie Cai}
\orcid{0000-0002-0582-555X}
\affiliation{%
  \institution{Tsinghua University}
  \city{Beijing}
  \country{China}
}
\email{jie-cai@mail.tsinghua.edu.cn}

\author{John M. Carroll}
\orcid{0000-0001-5189-337X}
\affiliation{%
  \institution{Pennsylvania State University}
  \city{University Park}
  \state{Pennsylvania}
  \country{USA}
}
\email{jmcarroll@psu.edu}

\begin{abstract}
 Infrastructure is an indispensable part of human life. In the past decades, the Human-Computer Interaction (HCI) community has paid increasing attention to human interactions with infrastructure. In this paper, we conducted a systematic literature review on infrastructure studies in SIGCHI, one of the most influential communities in HCI. We collected a total of 190 primary studies; the corpus includes studies published between 2006 and 2024. Most of the studies are inspired by Susan Leigh Star's notion of infrastructure. We discover three themes of infrastructure studies, including growing infrastructure, appropriating infrastructure, and coping with infrastructure. We foreground the overall trend of infrastructure studies in SIGCHI, which focuses on informal infrastructural activities in various socio-technical contexts. Especially, we discuss studies that problematize infrastructures and alert the HCI community about the underlying harmful side of infrastructure. 

\end{abstract}



\maketitle

\section{Introduction}
\label{sec:Introduction}

Infrastructures are large-scale systems that support people’s routines, like power grids that ensure residents’ electricity supply. In their seminal work “Steps Toward an Ecology of Infrastructure: Design and Access for Large Information Spaces” \cite{Star1996}, Susan Leigh Star and Karen Ruhleder reconsidered the term \textit{infrastructure} as a conceptual framework for information system studies. They proposed that infrastructure is relational and socio-technical. Infrastructure includes not only technical and physical foundations (e.g. generator stations, transmission lines, and towers) but also the actors and activities around the foundations (e.g. engineers and staff and their maintenance work). Specifically, Star and Ruhleder further elaborated the concept of infrastructure through eight characteristics \cite{Star1996}: 1) infrastructure is embedded into its environment; 2) infrastructure is transparent when at work; 3) infrastructure can reach beyond a single event or site; 4) new participants learn infrastructure through becoming a member of it; 5) infrastructure shapes and is shaped by conventions; 6) infrastructure is plugged into other infrastructures in a standardized manner; 7) infrastructure is built on installed bases; 8) infrastructure becomes visible upon breakdown.

In the recent three decades, infrastructure has become an increasingly prevalent framework in various fields related to HCI, like Science, Technology and Society (STS) \cite{LeDantec2013,Vertesi2014}, Computer-Supported Cooperative Work (CSCW) \cite{Karasti2018,Pipek2017}, Participatory Design (PD) \cite{Karasti2014,Neumann1996,Karasti2004,Clement2012}, Ubiquitous Computing (UbiComp) \cite{10.1145/2493432.2493497,Mainwaring2004}, Information \& Communication Technologies and Development (ICTD) \cite{Sambasivan2010,Pal2016}, and Accessible Computing \cite{Pal2016,10.1145/2049536.2049542,10.1145/3396076}. On the one hand, scholars use infrastructure to comprehend large-scale and distributed systems in different contexts \cite{Bowker1996,Bowker2010,Hanseth1996,10.1177/0165551509336705,Atkins2003,Hey2005,Ribes2010}. For instance, Accessible Computing researchers \cite{Pal2016} use \textit{accessible infrastructure} to describe "the social, economic, and technical conditions that form the larger environment in which accessible technologies are usable. \cite{Pal2016}". CSCW researchers refer to networked information technologies supporting scientiﬁc research activities as \textit{cyberinfrastructure} \cite{Atkins2003,Hey2005,Ribes2010}. STS researchers combine the concept of infrastructure with currently popular \textit{platform studies} and argue that "platform-based services acquire characteristics of infrastructure, while both new and existing infrastructures are built or reorganized on the logic of platforms" \cite{Plantin2018}. On the other hand, scholars explore methodologies that could better comprehend human-infrastructure interactions. Star introduced ethnography methods to investigate infrastructures \cite{Star1999}. Participatory Design researchers \cite{Bjorgvinsson2010,Ehn2008,Karasti2004} use \textit{infrastructuring} as an approach to encourage users’ engagement in infrastructure design, while Information Scientists \cite{Pipek2009} develop \textit{infrastructuring} as an analytical unit that indicates users work for making an infrastructure work. STS scholar Vertesi \cite{Vertesi2014}, who was inspired by UbiComp scholar Weiser \cite{Weiser1994,Weiser1993} and Chalmers \cite{Chalmers2004}, proposed \textit{seamful spaces} to research the heterogeneity in multi-infrastructure environments.

Additionally, the popularity of infrastructure causes discipline-level efforts to review its impact and set agendas for future research. Scholars in various fields conduct literature reviews of infrastructure studies \cite{Henningsson2017,Karasti2014,Korn2015,Young2017}. For example, Helena Karasti \cite{Karasti2014} traced how the PD community adopted and developed the concept of \textit{infrastructuring}. By reviewing the literature on infrastructuring in participatory design, Karasti emphasized the social and relational side of information infrastructure; in addition, Karasti also discusses how infrastructure can scale beyond a community by constructing the field with reflexivity and by understanding the political nature of people. This helped us better recognize the importance of social factors in infrastructure studies when doing our review. Besides, Inman and Ribes \cite{10.1145/3290605.3300508} reviewed seamless design and seamful design, two concepts that are closely relevant to infrastructure studies, in Ubiquitous Computing literature. They discussed the contexts and situations where designs should be seamful or seamless. The discussion of seam also inspired our discussion of transparency/visibility and problematization of infrastructure (more details in the discussion section).

The work mentioned above in various domains proves that infrastructure studies have great value in understanding the design, management, and use of large-scale systems. Such value of infrastructure studies is also applicable to HCI, where people focus on interactions between humans and technology. Moreover, infrastructure studies also have been increasingly prevalent in HCI \cite{10.1145/3025453.3025959,Lee2006,Zhou2020}. The value and popularity of infrastructure studies in HCI warrant a review. That said, little research has been done to systematically review how this concept was adopted, developed, and internalized by HCI people. \textbf{To fill the gap, we present a literature review of infrastructure studies in HCI.} Specifically, we chose infrastructure studies in SIGCHI as the sample. SIGCHI is one of the most influential communities in the field of HCI and one of the most popular databases of literature reviews in HCI \cite{Nelimarkka2019,Bopp2020,NunesVilaza2022}; therefore, infrastructure studies in SIGCHI could be representative of research on infrastructure in HCI.

We take a systematic approach to conducting the literature review. We collected a total of 190 infrastructure studies. From the 190 studies, we observe that from 2006 to 2024, the SIGCHI community had been increasingly interested in infrastructure studies. In addition, most of the infrastructure studies in SIGCHI (143 out of 190) adopted Star’s notion of infrastructure \cite{Star1996,Star1999,Star2006,Bowker1999,Neumann1996} as a theoretical foundation. We apply thematic analysis \cite{Braun2006b,Fereday2006} to capture the themes among the infrastructure studies in HCI. Three themes emerge from the systematic review, including growing, appropriating, and coping with infrastructure. Growing infrastructure includes developing, sustaining, and repairing infrastructures. Appropriating infrastructure describes how people utilize infrastructures to facilitate collaboration and participation. Coping with infrastructure shows interest in situations where infrastructures constrain or fail to support people; people in such situations have to come up with countermeasures to overcome the adversities caused by infrastructures. To be noted, by reviewing the papers, we also identify the dynamics of infrastructure stakeholders. Infrastructure studies cover a wide range of stakeholders. Infrastructure studies in anthropology \cite{Larkin2013}, STS \cite{barry2020material}, and geography \cite{furlong_small_2011}, have emphasized that infrastructure stakeholders' roles and characteristics are situational and influenced by various factors. That said, we wanted to foreground a pattern throughout the dossier of infrastructure studies we reviewed: there exists a significant discrepancy in terms of power, knowledge, and resources among different infrastructure stakeholders in infrastructure studies. And the different status would influence their experiences with infrastructure and even the way researchers study the experiences. We elaborate on this part in the findings section and conclusion section.

Our contributions to the HCI literature are threefold. First, we document the corpus of infrastructure studies in SIGCHI with a clear, thorough, and justified methodology. Second, from the systematic review, we discover three themes of infrastructure studies, including growing, appropriating, and coping with infrastructure. Third, with the findings, we discuss studying HCI from an infrastructure perspective, and the problematization of infrastructure; we also provide the epistemological and methodological implications for infrastructure studies in HCI based on the discussion.

\section{Methodology}

\label{sec:Methodology}
We took a systematic approach to reviewing the literature on infrastructure studies in HCI. We selected SIGCHI publications as the database of HCI studies. SIGCHI is one of the most influential communities in HCI and it publishes valuable and representative HCI literature. A number of HCI literature reviews have used SIGCHI as the primary data source to collect HCI studies \cite{Nelimarkka2019,Bopp2020,NunesVilaza2022}. In addition, a systematic literature review requires reviewers to state and document the methodology clearly and thoroughly \cite{Khan2003,Kitchenham2004,Pittaway2011}. Thus, we elaborated the entire procedure step by step. 

To be noted, we aimed at collecting and analyzing papers that explicitly discussed infrastructure as a specific \textbf{socio-technical} concept (i.e. providing a definition of infrastructure that involves such nature in the paper), and primarily investigated the interaction between infrastructure and some population. An example is Semaan et al.'s work “Transition Resilience with ICTs: ‘Identity Awareness’ in Veteran Re-integration \cite{10.1145/2858036.2858109}.” When veterans returned to civil society after military service, some of them experienced problems like post-traumatic stress disorder (PTSD). The study introduced a social support infrastructure for such veterans’ life transitions. The authors defined the infrastructure as “patterns of social connections and relationships among people enacted through various ICT-mediated and offline networks. \cite{10.1145/2858036.2858109}” The infrastructure aligned the interactions in support of the goal of helping veterans. The study investigated the interaction between veterans and the social support infrastructure. The investigation revealed that the formal social support infrastructure provided by governments was not capable of helping veterans. In facing the collapse of the formal infrastructure, veterans constructed their own social support infrastructure with ICT.

\subsection{Data Collection}

\begin{figure}[htp]
    \centering
    \includegraphics[width=0.8\textwidth]{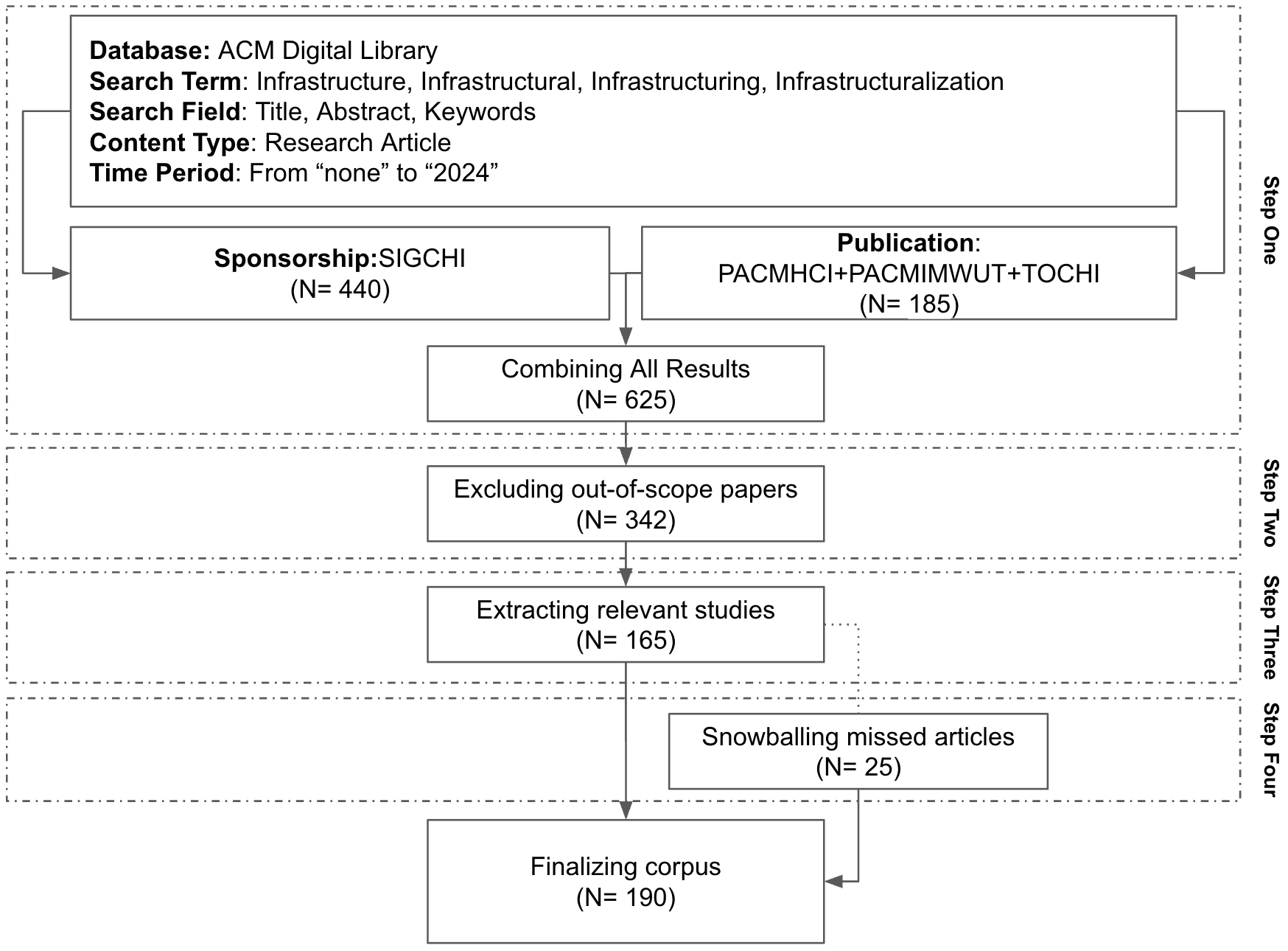}
    \caption{The flowchart of the data collection process.}
    \label{Fig:flowchart}
\end{figure}

The data collection consists of four steps (see Fig. \ref{Fig:flowchart}). \textbf{1)} We went to the ACM Digital Library to collect relevant primary studies. We searched for all papers published before 2024. In total, the search process returned 625 items eligible for review. \textbf{2)} Based on the criteria of previous HCI literature reviews, we excluded workshops \cite{VanMechelen2020,Salah2014,Bopp2020,Seering2018}, extended abstracts \cite{VanMechelen2020,Villarreal-Narvaez2020,Seering2018}, doctoral or master theses \cite{Villarreal-Narvaez2020,Baykal2020,Seering2018}, editorials \cite{VanMechelen2020}, and posters \cite{Salah2014}, because those publications are considered to lack detailed descriptions of research \cite{VanMechelen2020,Bopp2020,Salah2014} or to be under-reviewed \cite{Villarreal-Narvaez2020}; we also excluded non-English papers. After the exclusion, 342 remained. \textbf{3)} We then extracted studies that explicitly discussed infrastructure as a \textbf{socio-technical} concept (detailed criteria mentioned earlier in this section). After step 3, 165 papers remained. \textbf{4)} We took an extra step to search for studies that could be missed in the former steps. We applied Wohlin’s snowballing methods \cite{Wohlin2014} for the additional search. When reviewing snowballed papers, we applied the same criteria used in previous steps for screening, which means all papers we collected from this step were also published by SIGCHI-sponsored venues, strongly reviewed, and closely relevant to infrastructure and HCI. The snowballing process added 25 articles. Therefore, we collected a total of 190 (N=165+25) primary studies (See in Appendix ~\ref{appdendix:list}).

\subsection{Data Analysis}

\begin{figure}[htp]
    \centering
    \includegraphics[width=\textwidth]{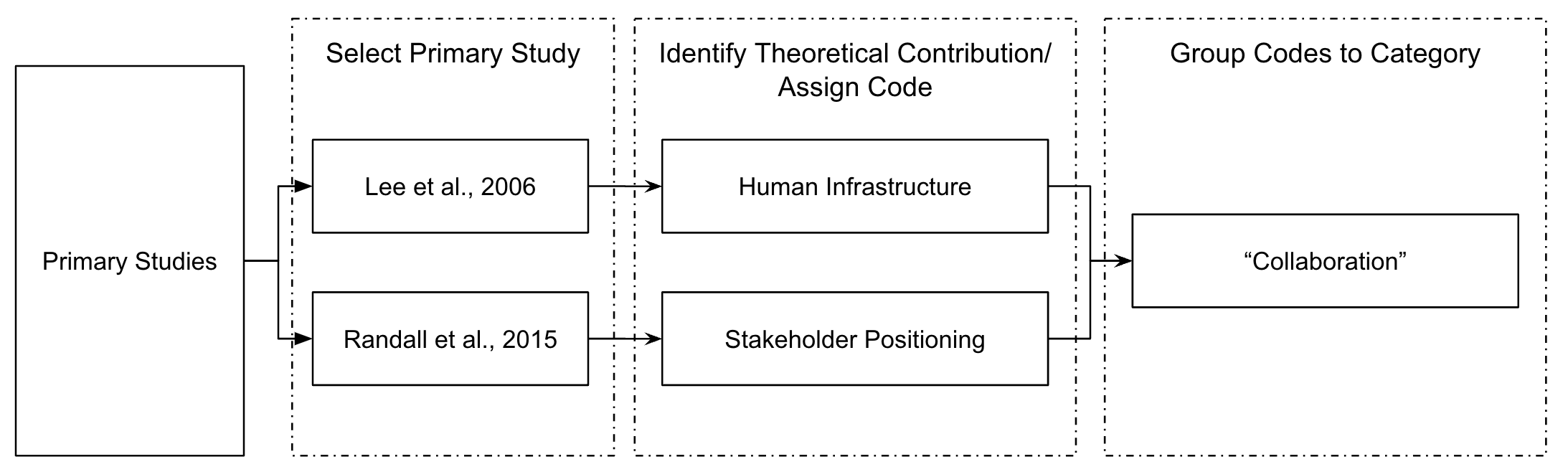}
    \caption{The flowchart of the data collection process.}
    \label{Fig:coding}
\end{figure}

We analyzed salient themes in infrastructure studies within HCI using thematic analysis \cite{Braun2006b,Fereday2006} to interpret primary studies. Many reviews examine the adoption of concepts in HCI \cite{Seering2018,Dillahunt2017}, often focusing on conceptual and theoretical innovations \cite{Seering2018}. Following this approach, we employed an inductive method to identify each paper’s theoretical foundation of infrastructure (how it conceptualized infrastructure) and its contributions to the literature. For example, Lee \cite{Lee2006} applied Star’s notion of infrastructure to human organizations and introduced \textit{human infrastructure} to describe labor organization supporting infrastructure. Such conceptual contributions formed the core of our thematic analysis. We collected and coded all theoretical contributions based on their specific aspects of infrastructure. Based on the coding results, we grouped codes that focused on similar aspects. For example, we found that Lee's \cite{Lee2006} \textit{human infrastructure} and Randall's \cite{10.1145/2702123.2702216} \textit{stakeholder positioning} both were about collaboration on utilizing an infrastructure for better efficiency. Therefore, we group these two codes, as well as other codes about collaboration, into one category "collaboration"  (See Fig \ref{Fig:coding}). Thereafter, we compared each code, group, and category back and forth to generate and refine themes.  

That said, due to the complexity of conceptual and theoretical innovation, some codes might cover multiple aspects of contributions to the infrastructure literature. Therefore, one single code could be used in multiple groups or themes. Finally, we identified three overarching themes: growing, appropriating, and coping with infrastructure. The themes are demonstrated in detail in Section ~\ref{sec:Themes}. In a sense, the three themes can be generally understood as three stages of the infrastructure lifecycle. Especially, the themes are mutually informed because of their close relationships. For instance, growing infrastructures involves the collaborative work of participants. In addition, when infrastructure breaks down and fails to provide support, participants have to come up with solutions to cope with the breakdown. Furthermore, given that some studies involve multiple research interests, they may belong to multiple themes at the same time. Thus, the total number of studies on three themes may not match the number of primary studies.

\section{Descriptive Statistics}
\label{sec:stats}
To provide a holistic overview of infrastructure studies in SIGCHI, we present the studies’ descriptive statistics, including publication venues and methods. The overview demonstrates that, from 2006 to 2023, the SIGCHI community had been increasingly interested in infrastructure studies; most infrastructure studies in SIGCHI applied qualitative methodologies.

\begin{figure}[htp]
    \centering
    \includegraphics[width=\textwidth]{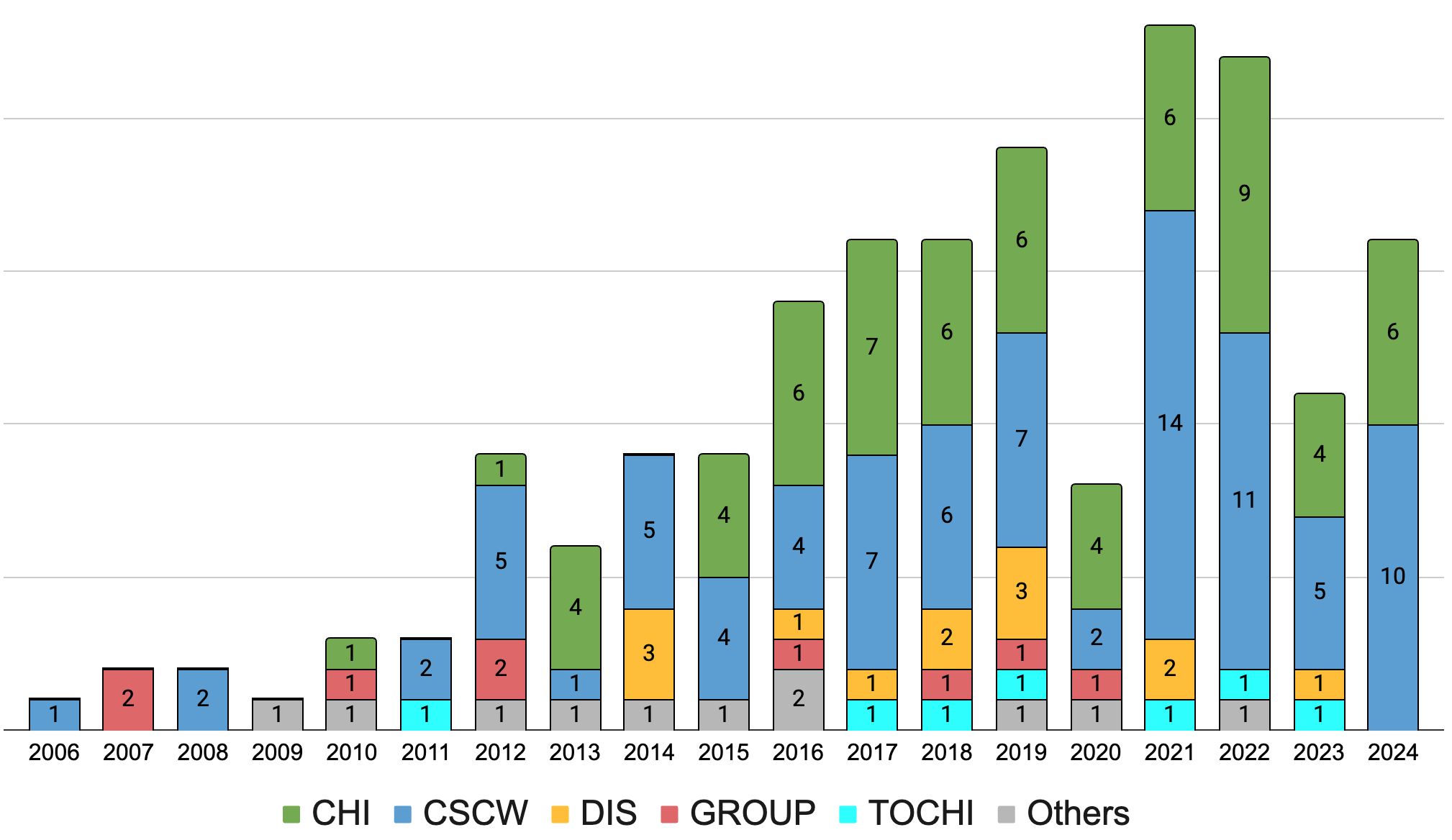}
    \caption{Publication venues of infrastructure studies in SIGCHI (X-axis for year, Y-axis for number of studies).}
    \label{fig:publication}
\end{figure}

\textbf{Publication Venues}. The study pool has a total of 190 primary studies. Most studies came from CSCW (N=86), CHI (N=64), DIS (N=13), GROUP (N=9), and TOCHI (N=6). Papers from these five venues constituted more than 90 percent of the entire pool (178 out of 190). A detailed description of the publication venues and publication years of all primary studies is displayed in Appendix ~\ref{appdendix:list}. We presented the distribution of publications by year in Fig. \ref{fig:publication}. To focus on the most popular publications, we only specify the names of the top five venues. The rest of the papers (including publications on UbiComp \cite{10.1145/1620545.1620570}, UIST \cite{Mortier2012}, MobileHCI \cite{10.1145/2785830.2785864}, CHI PLAY \cite{10.1145/3410404.3414259}, CABS \cite{10.1145/2631488.2631500}, etc.) are labeled as "Others." From the statistics results and the distribution diagram, we conclude that infrastructure studies are becoming more and more popular in SIGCHI. From 2006 to now, at least one infrastructure study has been published in SIGCHI every year. Especially, the number of papers significantly increased after 2011. In 2006-2011, SIGCHI researchers produced less than 2 papers per year on average; in 2012-2024, they yielded more than 13 papers each year. The increasing number of publications each year indicates that the SIGCHI community’s interest in infrastructure studies has been steadily growing. 

\textbf{Research Methods}. Among all primary studies (N=190), 174 took qualitative approaches; 16 applied mixed methods. In terms of data collection methods, 170 of them deployed interviews, either independently or combined with other methods like observation or a focus group; 6 analyzed archived documents; 3 collected data through workshops; 3 used case studies; 5 collected other types of data like online comments or participants’ reflection; there were also 3 studies that did not demonstrate their data collection methods. Regarding data analysis, 61 employed grounded theory; 51 used thematic analysis; 16 applied inductive coding; 8 deployed deductive coding; 9 analyzed through a combination of inductive and deductive coding; 15 mentioned other qualitative methods like affinity analysis, interpretative analysis, or critical reading; 30 studies did not specify their data analysis methods.

\textbf{Concluding Notes}. The rise of infrastructure studies within SIGCHI reflects a growing recognition of theories that emphasize social factors and large-scale systems in HCI. Initially, since its inception around the 1940s, HCI predominantly focused on technical problems, with research objectives centered around personnel or staff who used technology systems in research or military institutions \cite{grudin_tool_2022}. Theoretical approaches in HCI during this period primarily employed cognitive theories to address HCI issues \cite{rogers_hci_2012}. However, in the late 2000s, with the widespread accessibility of social media and smartphones, the landscape began to shift. The user base (the "H" in HCI) diversified significantly. The technology component (the "C" in HCI) expanded to encompass more than just a single device, evolving into a complex amalgamation of multiple information systems. Concurrently, the nature of the interaction (the "I" in HCI) became more contextual and situational \cite{grudin_tool_2022}. This period marked a paradigm shift in HCI, with a heightened focus on social factors. Yvonne Rogers, for instance, referred to HCI theories during this time as "contemporary HCI," highlighting "the emergence of a more self-conscious reflexivity and social conscience" \cite{rogers_hci_2012}. John M. Carroll characterized this era's focus as the "socially and materially embedded experience of users" \cite{carroll2013human}. Therefore, it was a natural progression for the concept of infrastructure, which not only underscores the social aspects of technology but also concentrates on large-scale systems, to become an increasingly prominent concept in the field.

Furthermore, the descriptive statistics provide a comprehensive understanding of infrastructure studies in SIGCHI. These studies have emerged as trending topics within the HCI community. For example, in 2021, SIGCHI published 23 papers, of which 7 addressed infrastructure breakdowns during the COVID-19 pandemic (e.g., \cite{Li2021,Cai2021}). This trend also mirrors the methodological evolution in HCI. The past decades have witnessed the emergence of what is known as "the third wave of HCI" \cite{Bodker2006,Duarte2016,Harrison2007}, advocating for contextual, situated, and nuanced interpretations of human-technology interactions. Correspondingly, infrastructure studies gained popularity in HCI during these decades. These studies often focus on specific situations where practices are deliberately organized to achieve particular objectives, demonstrating an interest in the intricacies of human-infrastructure interactions. This historical context has significantly influenced the paradigm of infrastructure studies. For instance, in terms of methodology, the majority of infrastructure studies utilize qualitative methods, such as interviews, which are adept at capturing the subtleties of human interactions with infrastructure, as opposed to quantitative approaches.

\section{Findings}
\label{sec:Themes}
We obtained three themes from the primary studies. The themes demonstrate the various interactions of actors with infrastructures, including growing, appropriating, and coping with infrastructure. Growing infrastructure includes initiating infrastructure development, sustaining infrastructure, and managing the repair of infrastructures. Appropriating infrastructure talks about utilizing infrastructures by emphasizing collaboration and promoting participation. Coping with infrastructure pays attention to situations where infrastructures constrained or failed to support people; people in such situations had to overcome the adversities caused by infrastructures.

\subsection{Growing Infrastructure}
We identified 64 articles that worked on growing infrastructure. Growing infrastructure refers to the activities that build and maintain infrastructures. Specifically, we introduce studies on how to initiate infrastructure development, how to sustain infrastructures, and how to manage infrastructures' repair work.

\subsubsection{\textbf{Initiating the Development}}

Infrastructures are large-scale and distributed systems, and such extensiveness is not achieved instantly. Rather, it takes a long process of infrastructural development. Besides, infrastructures are built based on installed bases, indicating that developing a new infrastructure involves numerous strategies to purposefully utilize existing infrastructures \cite{Star1996}. 

The most salient topic on initiating infrastructure development was leveraging participants' speculations of future infrastructures \cite{Soden2019,10.1145/3313831.3376515,10.1145/2470654.2466152,10.1145/3290605.3300496,10.1145/2675133.2675298}. Researchers were aware of the importance of various speculations in initiating an infrastructure. They created numerous concepts to demonstrate the speculations \cite{10.1145/3313831.3376515,10.1145/2470654.2466152,Soden2019,10.1145/2675133.2675298}, which aimed at understanding participants' expectations for potential infrastructure. One exemplar is Wong's definition of \textit{infrastructural speculation} \cite{10.1145/3313831.3376515}. Wong and colleagues defined infrastructural speculation as "an orientation towards speculation that aims to interrogate and ask questions about the broader lifeworld within which speculative artifacts sit, placing the lifeworld (rather than an individual artifact) at the center of a designer’s concern" \cite{10.1145/3313831.3376515}. They drew particular attention to \textit{lifeworld}, which emphasized, "the things that must be true, common-sense, and taken-for-granted in order for the design to work" \cite{10.1145/3313831.3376515}. The emphasis on the importance of 'lifeworld' when developing 'infrastructural speculation' shows infrastructure researchers' concern with the relationship between the future infrastructure and the current infrastructures. 

This topic is also consistent with ICTD literature on people's motivation to build new infrastructure. Hussain et al. \cite{10.1145/3392561.3394640} reported a study of Rohingya refugees in Bangladesh, who were suffering from the lack of infrastructure yet still desired to build one for a better future. The authors foregrounded \textit{social hope}, a unique type of speculation among refugees, that focused on "a practical and feasible future that is rooted in the morality of a community \cite{10.1145/3392561.3394640}" as the key motivation of infrastructure development.

\subsubsection{\textbf{Managing Endurance}}
\label{subsec:maintenance}

After infrastructure is built, another topic emerges: how to manage its life cycle. As Participatory Design researchers \cite{Karasti2004b} emphasize, infrastructure should be considered "from one-time technology development towards ongoing processes" \cite{Karasti2004b}. The "ongoing process" point of view emphasized the need of sustaining infrastructure for the long term. In our study pool, we observed a group of researchers that worked on the sustainability of infrastructure \cite{10.1145/2531602.2531692,10.1145/1958824.1958861,10.1145/2531602.2531736,10.1145/2531602.2531700,10.1145/2702123.2702216,10.1145/2531602.2531624}. 

There were many concerns about the longevity of infrastructure. The most popular concern of sustainability was time management. Researchers wanted infrastructure to stay functional for a long period of time. Ribes and Finholt \cite{10.1145/1316624.1316659} pointed out that infrastructure is "changing at a rapid and ever-increasing pace; yesterday’s novel solutions quickly become today’s staple resources and even faster seem to become tomorrow’s relics." \cite{10.1145/1316624.1316659} The juxtaposition of yesterday, today, and tomorrow spoke to the inertia of installed base \cite{Star1996} and revealed the change of infrastructure as time goes by. Another concern was capacity management \cite{10.1145/1316624.1316660,10.1145/2531602.2531700,10.1145/2531602.2531624}. As infrastructure grows, the scale, personnel, and capability also change. The scale issues were due to the expanding coverage of the growing infrastructure. To better manage the scale of infrastructure, researchers proposed various attributes that could demonstrate the scale and capacity of infrastructure. Ribes \cite{10.1145/2531602.2531624} showed interests in "[infrastructure] actors’ techniques and technologies for knowing and managing large-scale enterprises, including the All-Hands Meeting, surveys, and descriptive statistics, and benchmark metrics – each of these scaling activities seeks to represent and manage the size and growth of these sociotechnical systems" \cite{10.1145/2531602.2531624}.

When an infrastructure malfunctions or fails, it breaks down and needs repair. Infrastructures work behind the scenes; the mechanism that determines the function of infrastructure remains invisible until the infrastructure breaks down \cite{Star1996}. Thus, breakdowns reversely foreground important elements which are oftentimes overlooked when infrastructures work smoothly \cite{10.1145/3322276.3322330,10.1145/2145204.2145319}. Especially in this theme, researchers found that infrastructure breakdown and repair not only reveal failures but also generated valuable insights for future design \cite{Houston2016,10.1145/2145204.2145335,10.1145/2858036.2858420}. Reflecting on this, researchers revisited the idea of repair of infrastructure from a long-term and positive perspective, considering it as a manageable property of infrastructure \cite{Jackson2014,10.1145/2531602.2531692,10.1145/2145204.2145224}. Like Rosner and Ames said \cite{10.1145/2531602.2531692}, "breakdown and repair are not processes that designers can effectively script ahead of time; instead, they emerge in everyday practice." The unpredictable and inevitable breakdowns and repair work formed a condition, where infrastructure participants had to constantly negotiate with the damage of breakdowns, the benefits of stakeholders, the efficacy of repair work, and potential risks \cite{10.1145/2531602.2531692}. 

Some studies also talked about how to end the life cycle of infrastructure \cite{10.1145/2818048.2820077}. Though it might take a long period of time, most infrastructures "die" someday. Considering the large scale of infrastructure, how to deal with the "death" of infrastructure is also an important research topic. Cohn et al. \cite{10.1145/2818048.2820077} reported the termination of an aging cyberinfrastructure, which was in the face of "hardware’s material decay, programming languages and software tools reaching the end of support, and obsolete managerial methodologies." The aging of the infrastructure involved a complex process where different components aged at different paces. Cohn stated that the termination of infrastructure also could be considered as a type of repair work that negotiated with the complexity of the aging process, while the aim of the repair work was not to maintain the infrastructure but to terminate it purposefully. Cohn called the "repair-into-decay" process as \textit{convivial decay}, showing a way that actively ends the life of infrastructure \cite{10.1145/2818048.2820077}.

\subsection{Appropriating Infrastructure}

We recognize 80 papers that investigate the appropriation of infrastructure. By appropriation, we mean the adoption and utilization of infrastructures to serve certain purposes. Studies of appropriating infrastructures include understanding collaboration and promoting participation. This theme describes the infrastructure from the perspective of users, who are supposed to leverage infrastructure for certain purposes: due to the heterogeneity of infrastructure, it requires multiple components' participation and collaboration so that it can work; therefore, facilitating participation and collaboration is the main focus of the theme.

\subsubsection{\textbf{Comprehending Collaboration}}

It is important to understand the importance of collaboration in appropriating infrastructure. Star and Ruhleder \cite{Star1996} emphasized that infrastructure components are not isolated, they are interconnected in various and complex ways. This also applies to the appropriation of infrastructure, i.e., making use of infrastructure. As Karasti \cite{Karasti2014} reiterated, making use of infrastructure always took place at a level higher than individual, such as community or society. Therefore, it is impossible for an individual to appropriate an infrastructure, it requires collaboration.

Collaboration in infrastructure is pervasive. It could be the cooperation between international companies \cite{10.1145/1841853.1841860}, the coordination between national policymakers and execution departments \cite{10.1145/2858036.2858543}, the connection of local education institutions \cite{10.1145/3310284,10.1145/3290605.3300729}, or communication among staff in the same workflow \cite{10.1145/1460563.1460657}. Collaboration makes huge differences in infrastructure appropriation, including reducing costs, saving time, and improving transparency \cite{10.1145/2631488.2631500}. Therefore, comprehending collaboration helped participants better appropriate infrastructure. There was a group of scholars who particularly investigated collaboration in infrastructure \cite{Lee2006,10.1145/3148330.3148345,10.1145/2493432.2493497,10.1145/1460563.1460581,10.1145/2998181.2998344,10.1145/3313831.3376658,10.1145/3290605.3300716,10.1145/3025453.3025959}.

One strand of the research was interested in the organization form of collaboration. Researchers proposed various frameworks to understand such organization \cite{Lee2006,10.1145/3328020.3353932}. The most salient contribution was the conception of \textit{human infrastructure} by Lee et al. \cite{Lee2006}. According to Lee, human infrastructure was "the arrangements of organizations and actors that must be brought into alignment in order for work to be accomplished" \cite{Lee2006}. Inspired by Star's notion of infrastructure, human infrastructure also viewed collaboration from an infrastructure lens, pointing out "a new way to understand organizational work, in contrast to traditional organizational structures, distributed teams, or networks" \cite{Lee2006}. Though the concept also covers aspects related to the development of infrastructure, it primarily talks about a novel form of human collaboration that supports an infrastructure. Therefore, we consider this as an example of collaboration. After its inception, human infrastructure soon became a popular lens for understanding collaboration in HCI \cite{10.1145/3491102.3517572}. It also laid a solid ground for later studies on human infrastructure in ICTD contexts \cite{Sambasivan2010,10.1145/3173574.3173958,10.1145/3173574.3174213,10.1145/3313831.3376658}. For instance, ICTD researchers Sambasivan and Smyth \cite{Sambasivan2010} reported that, in rural areas, human infrastructure played a significant role in helping local residents overcome the constraints due to the lack of technology resources. Specifically, they emphasized that the human infrastructure, which consisted of local residents and their social networks, was pervasive, flexible, low-cost, and robust.

Another strand analyzed collaboration among infrastructure components, primarily from a relational perspective \cite{Agid2016,10.1145/2702123.2702216,10.1145/3148330.3148345}. Infrastructure is relational \cite{Star1996}: the successful utilization of infrastructure involves the careful arrangement of relationships among various components. The relational nature of infrastructure provided rich implications for research on collaboration work in infrastructure. The relationship that researchers investigated were various, including the relationship among collaborators in one work infrastructure \cite{10.1145/2702123.2702216} as well as the relationship between two infrastructures \cite{10.1145/3148330.3148345}.

\subsubsection{\textbf{Promoting Participation}}

The appropriation of infrastructure requires collective work, and collective work needs the participation of multiple participants. Therefore, while paying attention to collaboration, infrastructure researchers were also interested in actors’ participation \cite{Celina2016,10.1145/2818048.2820015,10.1145/3025453.3025606}. Scholars identified participation problems in infrastructures. Some demonstrated the lack of engagement of participants \cite{10.1145/3173574.3174081}. Taylor et al. \cite{10.1145/3173574.3174081} investigated a UK community’s use of local civic infrastructures. When residents in the community could not generate emotional attachment to technologies through participation, they lost interest in participating. Others also described low productivity issues \cite{10.1145/2389176.2389184,10.1145/2702123.2702508} due to frustration, fear of criticism, or conflicts among participants.   

Researchers then looked for solutions to the participation problems. One type of solution focused on acknowledging participants’ contributions. Infrastructures worked invisibly \cite{Star1996}, so the participants' contributions were also hidden. The invisibility oftentimes led to ignorance of contribution \cite{10.1145/2957276.2957295,Irani2013,Pendse2020}. To counter this, Bullard \cite{10.1145/2957276.2957295} introduced the motivation strategies in a volunteer-based website. Leaders of the website constructed a community exclusively for invisible contributors. The community of insiders certified and honored such contributors’ expertise and effort. Therefore, contributors were motivated to contribute more.

Another type of solution worked on creating a comfortable environment for participation. Scholars were aware of the importance of social factors in infrastructures \cite{Lee2019,10.1145/3025453.3025979,Ludwig2017,10.1145/2598510.2598528,10.1145/3196709.3196758,10.1145/2675133.2675153}. An environment that fostered social interactions among participants, as argued by Lee et al. \cite{Lee2019}, encouraged mutual support that motivated individual workers. On the contrary, a workplace, with social structures that only favored senior members, frustrated new participants and discouraged their participation \cite{10.1145/2389176.2389184}. In addition to social factors, some researchers paid attention to the infrastructural factors that influenced participation. Irani et al. \cite{10.1145/1841853.1841860} revealed that infrastructure could symbolize its users' characteristics. The characteristics, which were visible and easy to detect, gave new collaborators an impression of the infrastructure and influenced their motivation to participate. If the impression made participants "feel right," the participation was promoted. To create a better environment, researchers also proposed various design methods \cite{10.1145/3196709.3196758,10.1145/1460563.1460581}. Green et al. \cite{10.1145/3025453.3025606} proposed to decentralize the power of infrastructure and allowed all participants to "collectively define the role and form" when designing and assigning tasks \cite{10.1145/3025453.3025606}.

\subsection{Coping with Infrastructure}
In the last two subsections, we introduced two themes focusing on how infrastructure could better support human activities. However, we also obtained 71 studies that showed that infrastructures are adverse rather than helpful to users. In “Infrastructure Problems in HCI,” Edwards et al. \cite{10.1145/1753326.1753390} listed three infrastructure problems against users: 1) users’ experiences could be constrained by designers’ decision-making; 2) users’ understanding of and collaboration around infrastructure could be hindered by the relational and contextual nature of infrastructure; 3) users’ interaction could lack support from infrastructure due to poor design. In this theme, we show particular interest in studies that viewed infrastructure from a negative perspective: that is to say, the users do not benefit from infrastructure: rather, they are constrained by it and have to conduct various work to overcome the constraints. Most studies on this theme built their work on Star's conception of infrastructure. For example, Star emphasized that infrastructure is learned through membership, which means that infrastructure prioritizes its members over others; according to studies in the current theme, such prioritization can lead to marginalization, discrimination, or exclusion and therefore constrain certain user groups.  

\subsubsection{\textbf{Encountering Constraints}}

Infrastructures, like a double-edged sword, can both support and constrain human activities \cite{10.1145/1880071.1880101,10.1145/1958824.1958881}. These constraints often negatively shape users' experiences. Wang et al. \cite{10.1145/2145204.2145294} observed that user experiences can be limited by various factors such as technical design, social environment, and physical arrangement of the infrastructure. It's important to note that this subtheme, focusing on infrastructure constraints, is distinct from the "Promoting Participation" subtheme. While "Promoting Participation" addresses scenarios where participants choose not to adopt an infrastructure due to internal reasons like lack of motivation or interest, the constraints subtheme highlights cases where users are motivated to use an infrastructure but are impeded by external factors. In essence, the former subtheme deals with internal barriers to adoption, whereas the latter focuses on external impediments to effective use.

Scholars reported four categories of constraints: infrastructural restrictions, infrastructural exclusion, unbalanced power distribution among infrastructure stakeholders, and nontransparent infrastructure mechanisms. Infrastructures have restrictions that restrict users' experiences \cite{Lee2019,Duarte2021,10.1145/3196709.3196749,10.1145/3025453.3025959,10.1145/2998181.2998290,10.1145/2818048.2820015,10.1145/3025453.3025889}. The restrictions could be intentionally set to regulate user behaviors. For example, Boustani et al. \cite{10.1145/3410404.3414259} documented that, online gamers often encountered restrictions from the infrastructure when they created usernames. The restrictions were determined by technology considerations (e.g. unique name for identification) or policies (e.g. content moderation). Researchers also reported participants' circumvention in the face of such limits. To bypass infrastructural constraints, "[users] wrestle against the affordances of the installed base of [an] Infrastructure, and take the shape of engaging or circumventing activities" \cite{10.1145/3025453.3025959}. The key point is to understand the installed base of infrastructure. To recognize the work and capability of understanding the current infrastructure and bypassing infrastructural constraints, Erickson and Jarrahi \cite{10.1145/2818048.2820015} defined the knowledge on such circumventing activities as \textit{infrastructure competence}.  

The study pool revealed exclusion issues as well \cite{Singh2017,10.1145/2598510.2598553}. Infrastructure could be learned by becoming a member of it \cite{Star1996}, indicating that access to infrastructure is dependent on membership. When designers blueprint infrastructures, they sometimes do not fully consider the diversity of user groups, like people with relatively low knowledge \cite{Sabie2019,Shen2020,Singh2017,10.1145/3359270} or disabilities \cite{10.1145/3313831.3376658}. Therefore, when the infrastructures are built, they only support some populations while leaving others excluded \cite{10.1145/2598510.2598553}. The exclusion could lead to bias \cite{Soden2019} and discrimination \cite{10.1145/2531602.2531687}. Feinberg et al. \cite{10.1145/2598510.2598553} stated that, infrastructural exclusion "represents the persistent vagueness, ambiguity, and invisibility that standard classificatory practice attempts to eliminate via the systematic application of technical rules to establish neatly differentiated relationships" \cite{10.1145/2598510.2598553}. Users had to come up with solutions when they were excluded. The primary solution was to draw attention to such excluded groups and hear their voices. For instance, Rajapakse et al. \cite{10.1145/3196709.3196749} introduced the lack of infrastructural support experienced by people with disabilities. The research group also proposed design artifacts to help people with disabilities to express their needs and assemble support from different societal resources.

In addition, some researchers pointed out the unbalanced power distribution in infrastructures \cite{10.1145/3173574.3174055,10.1145/3290605.3300810,Li2021}. One infrastructure involves various stakeholders, and different groups of stakeholders might utilize the infrastructure in different ways. When the power across groups is distributed unevenly, the infrastructure tends to prioritize the more powerful ones over others \cite{Soden2019}. The unevenly distributed power in the infrastructure could cause conflicts. Leal et al. \cite{10.1145/3290605.3300810} reported a case on the armed struggles of a Colombian guerrilla group with the Colombian army. The Colombian army, who had more resources, introduced technologies to locate the guerrilla group’s communication and then sent targeted bombs. The technologies and their threats created an infrastructure that had "a specific aim in view: the destabilisation and eventual destruction of an opposition" \cite{10.1145/3290605.3300810}.

There was also a line of research interested in nontransparent infrastructure mechanisms. Infrastructure works behind the scenes \cite{Star1996}. Ackerman et al. \cite{10.1145/1880071.1880101} introduced a code infrastructure of an online community. It invisibly surveilled users’ activities and conducted immediate punishment when users misbehaved. The creepy nature of the infrastructure’s work caused trust issues \cite{10.1145/2470654.2466152}. To deal with such issues, users raised awareness of situations where infrastructures could conduct invisible work \cite{10.1145/2470654.2466467} and foregrounded the work in the black box for more transparency \cite{10.1145/2380116.2380143}.

\subsubsection{\textbf{Resolving Failures}}
\label{subsubsec: resol}

In this subsection, we explore the failures of infrastructure and the efforts made by participants to address these failures. It is important to note that while the scenarios of infrastructure failure discussed here bear similarities to the breakdowns covered in Section~\ref{subsec:maintenance}, our focus in this section is specifically on the impact of these failures on participants and their strategies for overcoming them. Additionally, this subtheme contrasts with the scenarios described in the "appropriating infrastructure" section. While the "appropriating infrastructure" section primarily addresses concerns relevant to infrastructure management, the current subtheme concentrates on individual responses to infrastructure challenges. According to researchers, infrastructure failures took place in a large variety of scenarios. Some failures were about the deconstruction of habitats due to military activities \cite{Ahmed2015,Semaan2011}, like the breakdowns of transportation, education, and power infrastructures in a war \cite{Semaan2011}. Some failures described the shortage of critical resources \cite{10.1145/2998181.2998290,10.1145/3375183,10.1145/2441776.2441783}. Dailey and Starbird \cite{10.1145/2998181.2998290} described the cutoff of information about victims after a landslide in the USA. Some failures revealed the malfunction of essential service departments \cite{10.1145/3432946,10.1145/3290605.3300688,10.1145/2858036.2858109}. Veterans in the USA \cite{10.1145/2858036.2858109} criticized the collapse of formal infrastructures, which was supposed to facilitate their life transition from the military to the civil world. Such failures caused tremendous trouble to users, yet users still actively looked for solutions to remain resilient and robust \cite{10.1145/1620545.1620570}.

Researchers documented various efforts conducted by users to resolve infrastructure failures. On the one hand, users’ countermeasures covered different spatial levels. Many of the countermeasures were at the community level. As disruption to people’s life, infrastructure failures often affect a huge group of users simultaneously. Therefore, users cooperated with others who had similar experiences to solve the problems \cite{10.1145/3432946}. The cooperation highlighted social relationships within communities, such as shared identities between community members \cite{Britton2019,10.1145/2858036.2858109}. For instance, due to the absence of formal support for new mothers, they had to work together to create online social networks to meet the specific social, political, and medical needs of women \cite{Britton2019}. The social relationship also emphasized trust in social networks \cite{10.1145/3359175}. As Semaan and Mark \cite{Semaan2011} reported, the 2nd Gulf War destroyed most infrastructures that were reliable in peaceful times, and civilians looking for public services could be ransomed or killed by militia and insurgent groups. Therefore, civilians had to seek resources from close friends or strangers who had been carefully tested. Despite the prevalence of community-level work, few studies documented individual-level efforts. Gui and Chen \cite{10.1145/3290605.3300688} primarily documented how caregivers coped with breakdowns of healthcare infrastructures as individuals. The caregivers corrected omissions of staff, fixed the misalignment or non-alignment between organizations, and bypassed the infrastructures’ spatial, temporal, policy, and financial limits. 

On the other hand, the reactions also included different temporal levels. Some of the reactions focused on short-term goals as they sought expedient solutions; in most cases, users' strategies were “irregular, opportunistic, adaptable, responsive and decentralized \cite{10.1145/3173574.3174055}.” Moreover, some aimed at long-term countermeasures. Some participants suffered from ongoing disruptions in daily life, like the life transitions of new mothers \cite{Britton2019}. They formalized their solutions as routines to cope with ongoing disruptions \cite{10.1145/2935334.2935352,10.1145/2785830.2785864}. Semaan \cite{10.1145/3359175} described the “\textit{routine infrastructuring}” of war survivors, retired veterans, and LGBTQ people. Routine infrastructuring refers to people’s work of "building everyday resilience with technology" \cite{10.1145/3359175}. For example, war survivors constantly saved electricity through daily activities in case of a shortage of power; retired veterans regularly communicated with other veterans to retain the collectivist culture which they had been used to through military experiences; LGBTQ people habitually composed, read, and shared stories that helped them reconstruct their identities in online communities. The effort on self-help, which often involves the re-appropriation of currently available resources, would further create new infrastructures for themselves.

\section{Discussion}
In this work, we reviewed 190 SIGCHI studies that investigated infrastructure. From the dossier, we obtained three salient themes. The themes present that infrastructure studies mainly focus on how people grow, appropriate, and cope with infrastructure. To summarize, we obtain three themes from the systematic review of infrastructure studies in SIGCHI. Star's notion of infrastructure serves as a significant theoretical foundation for most researchers to understand and contribute to the infrastructure literature. In the following section, we will discuss how infrastructure studies emphasize informal infrastructural activities, as well as how HCI researchers problematize infrastructure.

\subsection{Infrastructure Studies in SIGCHI: A Focus on Informal Activities}

Infrastructure predicates an environment where human life is supported and shaped by the infrastructure. Infrastructure manifests as an ongoing process that needs to be negotiated and re-negotiated all the time \cite{10.1145/2531602.2531692}. Therefore, infrastructure is situated in dynamic human contexts; it must be constantly critiqued and reconsidered. Yet this requirement inherently conflicts with social inertial tendencies for practices and power structures to persist. In the findings, we identified a pattern that to better understand the relational nature of infrastructure, researchers need to have an awareness of informal activities. Many previous studies have discussed similar concepts in HCI or infrastructure studies. Bowker and Star pointed out the complexity of classification and standards in infrastructure that cause "the problem of residual categories" \cite{Bowker1999}. We list four significant types of informal activities across the papers reviewed.

First, the review highlights the human experience's \textit{dependence} on infrastructural change. The concept of human infrastructure \cite{Lee2006} illustrates this dependence by emphasizing that infrastructure is not just a technical system but also a social arrangement, one that relies on human organization and labor to function. This perspective foregrounds the "work of infrastructure" as an ongoing process rather than a static structure. Besides, as Star \cite{Star1996} points out, infrastructure is not a fixed entity; it continuously evolves through negotiation and renegotiation, shaped by shifting contexts. These changes are not always foreseeable at the time of infrastructure design but emerge dynamically over time, often through disruptions and adjustments. Because the alignment of infrastructural contexts is inherently situated and cannot be predetermined by rigid frameworks, researchers focusing on infrastructure sustainability advocate for the design of more flexible, adaptable, and open-ended infrastructures \cite{10.1145/2531602.2531692}.

Besides, the review draws attention to participants' \textit{amateur} work that overcomes infrastructural constraints, too. As we introduced in the descriptive statistics section, SIGCHI has been paying increasing attention to infrastructure in the Global South; among the studies in the Global South, a large proportion of them are concerned about conditions where participants have difficulty getting support from infrastructure in their daily lives due to rurality \cite{Duarte2021}. Associated with the lack of infrastructure support is the lack of systematic and professional expertise, which could help participants overcome the constraints. What participants can do is try to use their relatively low knowledge to solve the problems. Researchers are interested in amateur work that creatively solves problems in the infrastructure. For instance, Chandra's work \cite{10.1145/3025453.3025643} introduced local residents' human labor that utilized social relationships to overcome the limitations of the lack of technical infrastructure of communication. This problem-solving presents the power of the grassroots against the institutional, professional, and systematic power behind an infrastructure. HCI researchers have been looking for ways to better support amateur work. Hoare et al. \cite{10.1145/2556288.2557298} argued for helping amateurs build social networks with other amateur workers and professionals. This could be a future direction for researchers who investigate and support amateur work in infrastructure studies.

In addition, the review sheds light on \textit{improvisational} practices in the face of infrastructural crises. By infrastructural crises, we mean conditions where crises like war \cite{Semaan2011} or nature disasters \cite{10.1145/2998181.2998290} have destroyed the infrastructures that people rely on. Infrastructure has the quality of being taken for granted and being invisible \cite{Star1996}. Therefore, when infrastructures are destroyed, participants experience situations where they suddenly lose the foundations of daily life and urgently need to get the foundations back. The sudden change and the short time for response push participants to make quick decisions with little preparation. Participants can only improvise to look for stopgaps that could help them survive the infrastructural crises. As presented in the third theme, Semaan and Mark \cite{Semaan2011} reported a study on how residents in a war reacted to supply shortages by changing lifestyles so that they could survive. The victims' improvisations were informal as they reacted with little preparation. The informal activities can be an important analytical unit for crisis informatics researchers to understand human behaviors when infrastructure breaks down. In HCI, researchers also have been studying improvisation. For instance, Kang and Jackson presented multiple studies \cite{kang_intermodulation_2018,kang_tech-art-theory_2021} on how people improvise as art practices. While these studies provide a solid foundation for understanding improvisation's essential characteristics such as reflexivity, tension, and interdependence \cite{kang_intermodulation_2018}, improvisation in this literature review can expand the discussion by putting it in a new context. For example, Kang and Jackson \cite{kang_tech-art-theory_2021} argued for making safe spaces for improvisational learning. However, in the papers we reviewed, most of the improvisations were in situations where people experienced infrastructure breakdowns and did not have "safe space." Therefore, by introducing new contexts, which are much more intensive, we can understand improvisation in a more comprehensive way.

Lastly, the review also speaks to \textit{invisible} labor in work settings. Infrastructure works behind the scenes. Like Liu \cite{10.1145/2493432.2493497} revealed, a major outcome of infrastructure actors' work is the transparency of the infrastructure. Therefore, the actors' contribution to the infrastructure is oftentimes invisible. The invisibility further makes it hard for formal frameworks to identify and recognize the contributions of infrastructure workers. This raises serious ethical issues, especially in work settings. Reflecting on invisibility issues, researchers have proposed various approaches to honor invisible contributions \cite{Irani2013}. That said, we also found new understandings of invisible labor. By analyzing examples like domestic worker, Star and Strauss \cite{star_layers_1999} alerted that work is recognized by contextual indicators, and failing to identify the indicators would result in invisible work. While most studies paid attention to indicators like productivity, we also found novel indicators to recognize labor. Bullard \cite{10.1145/2957276.2957295} found that working behind the scenes with a small group of people actually generates a sense of being privileged. In this case, being not recognized generates a feeling of privilege. The contradictory findings suggest future studies on invisible work in infrastructure. To be noted, the study on invisibility is not to merely foreground the invisible work, but also to explore the situatedness of the invisible work. For instance, researchers can pay attention to looking for different contexts where people prefer (or do not prefer) invisible work and the motivations behind such preferences.  


\subsection{Problematizing Infrastructure: How Does Infrastructure Become Troublesome?}


In the three themes presented in the findings section, we illustrated various problems with infrastructure, ranging from how to sustain it to how to overcome its constraints. In this section, we would like to discuss how such studies in HCI problematize infrastructure. One type of infrastructure problem emphasizes the way contexts, resources, and work are aligned. As the "when of infrastructure" emphasizes the relational nature of infrastructure, it points out the importance of understanding and aligning contexts so that creative arrangements can be made. For instance, one strand of infrastructure research in HCI particularly considers the \textit{seamfulness} as the unit of analysis. STS researcher Janet Vertesi \cite{Vertesi2014} used \textit{seamfulness} to denote the heterogeneity within or across infrastructures \cite{10.1145/3290605.3300508}. The heterogeneity could be problems caused by different standards of power grid used in different countries. In our study pool, seamfulness manifests as the scattered sources of critical information; they are waiting for users in disasters to piece together \cite{10.1145/2998181.2998290}. The "seamfulness" lens has inspired a large number of infrastructure researchers, as it could explain infrastructure problems in multiple dimensions, like technological, physical, and social factors. Another type of problematization, on the other hand, focuses on the purpose of the infrastructure. In the study pool, we also hear a small yet vehement voice that criticizes infrastructures not because they fail to be helpful, but because they never meant to help. As presented in the third theme, we see infrastructuralized platforms that marginalized people with low tech literacy \cite{Shen2020}; we report infrastructural violence built to censor or suppress citizen online speeches \cite{Li2021}. Such concerns have been reported by not only HCI researchers in SIGCHI (our study pool) but also researchers in Accessible Computing, who argue for more inclusive and accessible infrastructure for people with disabilities \cite{10.1145/3396076,Pal2016,10.1145/2049536.2049542}. 

To understand the purposes of infrastructure, we also need to draw on Star's notion of infrastructure. Star pointed out that \textit{membership} is one of the keys to getting access to an infrastructure \cite{Star1996}. However, subsequent infrastructure studies have shown that membership incorporates conflicts in inclusivity: infrastructure for some members sometimes means it only serves its members but not others. Therefore, infrastructure becomes a burden rather than a foundation for people without membership. The studies presented in the findings, especially in the third theme, reveal that infrastructures can be problematic at the moment when they are conceptualized. According to STS researchers \cite{LeDantec2013}, the conceptualization of infrastructure is usually associated with the idea of \textit{public}, where a set of facilities or services is open-access to everyone. But if the infrastructure cannot ensure the principle of being public, it can only become a gated residential area rather than a commonplace. The issues engendered by the conflict between \textit{membership} and \textit{public} point to a complementary approach to the aforementioned lens in terms of problematizing infrastructure. This approach is more concerned with the value of infrastructure rather than the alignment of contexts. The value-embedded design has been prevalent in HCI \cite{Molich2001,Cockton2004} and now we bring the conversation into the infrastructure context. In the infrastructure context, value is created by humans and expressed in the form of infrastructure. An infrastructure's goal and its participants' benefits are at odds when the values of the infrastructure clash with the values of participants. But what is the "human" or "participant" in this context? Star has stated that in infrastructure, the roles of designer, user, and repairer are blurred \cite{Star1996}. Therefore, to make infrastructure more inclusive, an idea is to blur the boundary between \textit{membership} and \textit{public}. Methodologically, researchers can pay specific attention to value conflicts at the infrastructure level. Social justice issues caused by infrastructures like marginalization \cite{Shen2020} can be considered as analytical units to reflect infrastructure problems; the problems can be imbrication \cite{Moitra2021}, torque \cite{Singh2017}, residuality \cite{10.1145/2598510.2598553}, or uneven power distribution \cite{Soden2019}, as we presented in the last theme. To investigate and fix these problems, researchers can learn from infrastructure studies in the Participatory Design community \cite{Karasti2004}. The PD community uses participatory design to listen to more groups when designing infrastructure; this method also gives more participants the right to shape infrastructure during its development.

In addition to participatory design, we advocate for the increased use of ethnographic studies in infrastructure research. Star, in 1999, highlighted the necessity of an ethnographic sensibility in infrastructure studies, proposing that "people make meanings based on their circumstances, and that these meanings would be inscribed into their judgments about the built information environment" \cite{Star1999}. Our review of existing literature revealed several studies employing ethnography to investigate infrastructure issues. A notable example is the work of Semaan and Mark \cite{Semaan2011}, who conducted an ethnographic study on residents' responses to infrastructure breakdowns during the 2nd Gulf War. Their research focused on the residents' experiences and usage of infrastructure in wartime, adopting a bottom-up approach. Such ethnographic studies are crucial for gaining a deeper understanding of residents' perceptions of infrastructure, particularly in scenarios where infrastructure fails, such as in war-torn areas. This approach is instrumental in identifying key aspects of infrastructure, thereby enabling more efficient resolution of infrastructure failures. We observed that social dynamics often emerge as the focal point in these ethnographic studies of infrastructure. Essentially, when examining infrastructures through an ethnographic lens, researchers tend to prioritize social, political, and cultural factors over technical aspects. This approach aligns with infrastructure studies in other domains, such as Information and Communication Technologies for Development (ICTD). For instance, Sambasivan and Smyth \cite{Sambasivan2010} explored how local residents in an underdeveloped area in India leveraged social relationships to compensate for technological limitations. Similarly, Hussain et al. \cite{10.1145/3392561.3394640} examined the social hierarchy and political dynamics among Rohingya Refugees as they endeavored to rebuild their infrastructure using ICTs for daily life. These studies collectively underscore the significance of ethnography in infrastructure research. Ethnography, with its emphasis on understanding people as the central component of infrastructure, provides researchers with invaluable insights into the user experience and the societal context of infrastructure usage.

In conclusion, we encourage future infrastructure studies in SIGCHI to pay attention to infrastructural problems. We would like to alert infrastructure scholars that not all infrastructures are created with the purpose of benefiting all human beings; some of them are inherently biased. Researchers should constantly review and examine the purposes of infrastructure to make them inclusive, accessible, and beneficial to the public. This strand of research can be informed by social justice issues, like discrimination, embedded in ICT systems. And more inclusive design methodologies, like participatory design and ethnography, can be applied in the studies.

\section{LIMITATIONS}
While claiming this review's contributions to the infrastructure literature in HCI, we also acknowledge that this work has several limitations because we do not cover all infrastructure studies in HCI. The first one concerns the sampling database. In this project, we choose to focus on SIGCHI-sponsored publication venues for our literature review. The landscape of literature reviews in Human-Computer Interaction (HCI) is vast, and the criteria for what constitutes an HCI study vary widely. Broadly, there are two main approaches to identifying HCI studies. The first approach is based on publication sources. For instance, Dillahunt et al., \cite{Dillahunt2017} "coded publication venues as HCI if the proceeding or journal’s site stated that human interaction with computing systems was a primary interest of the venue." This method involves selecting papers from specific publication venues or publishers known for their HCI content. For example, some reviews concentrate on works from certain publishers \cite{Frich2018,AltarribaBertran2019,Terzimehic2019}; notably, Altarriba Bertran et al. \cite{AltarribaBertran2019} utilize the ACM Digital Library for sourcing HCI literature. Others base their selection on a collection of publications recognized by academic organizations \cite{Caraban2019,10.1145/3359298}; a case in point is Caraban et al. \cite{Caraban2019}, who select the top 15 HCI publication venues as listed by Google Scholar. Additionally, some reviews choose multiple publications united by common themes \cite{Sanches2019,Nelimarkka2019}; for instance, Nelimarkka \cite{Nelimarkka2019} identifies HCI papers published in venues sponsored by SIGCHI. The second approach is driven by keywords or content. Kannabiran et al. \cite{kannabiran_how_2011} define HCI studies as those "already published or available in a public domain and affiliate itself to HCI or its cognate fields." Bopp and Voida \cite{Bopp2020} adopt a more practical strategy, categorizing a study as HCI if it falls under HCI-related classifications in the ACM Computing Classification System (CCS). Sampling literature based on publishers or specific publications tends to be a popular method in HCI literature reviews \cite{NunesVilaza2022,Nelimarkka2019,stefanidi_literature_2023}. For example, Stefanidi et al. in their recent work "Literature Reviews in HCI: A Review of Reviews" presented at CHI'23, choose SIGCHI-sponsored venues as their data source for HCI reviews, justifying this choice by stating that "the SIGCHI conferences and the way they have been shaped in the last 40 years accurately describe the intellectual development of HCI" \cite{stefanidi_literature_2023}. We adopt a similar approach in our study by selecting SIGCHI-sponsored publications as our sample pool. However, it is not our intention to suggest that SIGCHI publications are the only or most suitable source for such a review. We recognize that different methodologies have their respective strengths and weaknesses. Our goal is to offer an overview of the various methods employed in previous literature reviews for identifying HCI studies. This summary is intended to assist future HCI literature reviews by presenting a spectrum of approaches used within the HCI community.

Another limitation concerns the keywords used in data collection. We are aware that, in addition to "infrastructure", infrastructure studies might also use other keywords such as "toolkit", "platform", etc. However, we still choose to focus on one keyword. The most important reason for this decision is because of the workload of screening. Researchers use words or phrases in various ways. When they use "toolkit", "system", "platform", or "infrastructure" in the title/abstract/keyword, it might not mean they consider them as specific concepts or theories. Therefore, considering too many keywords might distract our search process and result in too many irrelevant studies. For instance, if we use "infrastructure", the first round would return about 400 papers for screening. But if we changed to other keywords such as "platform", the search would result in about 1000 papers. This would significantly increase the workload of screening papers, especially when the screening process includes thoroughly reading each paper to identify the definition, application, and discussion of infrastructure (details in the data collection section). Finally, we choose to focus on one word and its variants.

We are aware that the current approach, which chooses to not cover certain studies, might lead to missing important relevant literature and harming our contribution. Therefore, to balance the feasibility and the comprehensiveness of the literature review, 1) we conduct a thorough round of forward and backward sampling after the search based on keywords; 2) we select the most representative and relevant infrastructure studies that we do not include in the sample but are closely relevant to the review; we introduce them as background knowledge of infrastructure or discuss them with our findings. This approach of making up for the missing papers not only helps us stay focused on infrastructure studies in SIGCHI, which already yield rich findings and implications, but also connects our discussion of infrastructure studies in SIGCHI to a broader level so that we do not ignore relevant studies in other fields.

\section{CONCLUSIONS}
In this paper, we present a systematic review of the literature on infrastructure studies in SIGCHI. We collected and analyzed 174 infrastructure studies. We discovered three themes that emerged from the analysis: sustaining, appropriating, and coping with infrastructure. In the first theme (growing infrastructure), infrastructure manifests as large systems that need extraordinary work to develop and maintain. In the second theme (appropriating infrastructure), the technical design of infrastructure serves for collaboration among individuals or organizations. How to navigate relationships (such as social relationships) among various organizations becomes the most important topic. The last theme (coping with infrastructure) represents a perspective that criticizes infrastructure. We also emphasized the influence and limits of Susan Leigh Star's work on HCI researchers' investigations of infrastructure. We discuss how infrastructure problems can be better framed in different settings. We alert that not all infrastructures are created to serve all populations; some of them have serious bias issues. We, therefore, recommend researchers use more inclusive design methodologies, like participatory design, when designing and developing infrastructures.  

\begin{acks}
Thanks for the reviewers’ comments.
\end{acks}

\bibliographystyle{ACM-Reference-Format}

\bibliography{infra_review,Zotero}

\appendix

\section{Appendix: The list of primary studies}
\label{appdendix:list}

\begin{center}
\begin{longtable}{|p{11cm}|p{1cm}|p{1cm}|}

\hline \multicolumn{1}{|l|}{\textbf{Title}} & \multicolumn{1}{|l|}{\textbf{Year}} & \multicolumn{1}{|l|}{\textbf{Publication}} \\ \hline 
\endfirsthead

\multicolumn{3}{c}%
{{\bfseries -- continued from previous page}} \\
\hline \multicolumn{1}{|l|}{\textbf{Title}} & \multicolumn{1}{|l|}{\textbf{Year}} & \multicolumn{1}{|l|}{\textbf{Publication}} \\ \hline 
\endhead

\hline \multicolumn{3}{|r|}{{Continued on next page}} \\ \hline
\endfoot

\hline 
\endlastfoot

The human infrastructure of cyberinfrastructure \cite{Lee2006}                                                                                         & 2006 & CSCW        \\ \hline
Tensions across the Scales: Planning Infrastructure for the Long-Term \cite{10.1145/1316624.1316659}                                                                    & 2007 & GROUP       \\\hline
Growing an Infrastructure: The Role of Gateway Organizations in Cultivating New Communities of Users \cite{10.1145/1316624.1316660}                                    & 2007 & GROUP       \\\hline
Colour management is a socio-technical problem \cite{10.1145/1460563.1460657}                                                                                          & 2008 & CSCW        \\\hline
Representing Community: Knowing Users in the Face of Changing Constituencies \cite{10.1145/1460563.1460581}                                                             & 2008 & CSCW        \\\hline
Ubicomp4D: Infrastructure and Interaction for International Development--the Case of Urban Indian Slums \cite{10.1145/1620545.1620570}                                 & 2009 & UbiComp     \\\hline
Shopping for Sharpies in Seattle: Mundane Infrastructures of Transnational Design \cite{10.1145/1841853.1841860}                                                       & 2010 & ICIC        \\\hline
Social Regulation in an Online Game: Uncovering the Problematics of Code \cite{10.1145/1880071.1880101}                                                                 & 2010 & GROUP       \\\hline
The Infrastructure Problem in HCI \cite{10.1145/1753326.1753390}                                                                                                        & 2010 & CHI         \\\hline
SELECT * FROM USER: Infrastructure and Socio-Technical Representation \cite{10.1145/1958824.1958881}                                                                                    & 2011 & CSCW        \\ \hline
Collaborative Rhythm: Temporal Dissonance and Alignment in Collaborative Scientific Work \cite{10.1145/1958824.1958861}     & 2011 & CSCW        \\ \hline
Technology-mediated social arrangements to resolve breakdowns in infrastructure during ongoing disruption \cite{Semaan2011} & 2011 & TOCHI \\ \hline
A Sociotechnical Exploration of Infrastructural Middleware Development \cite{10.1145/2145204.2145404}                                                                  & 2012 & CSCW        \\ \hline
The Sociality of Fieldwork: Designing for Social Science Research Practice and Collaboration \cite{10.1145/2389176.2389183}                                            & 2012 & GROUP       \\ \hline
Sustaining the Development of Cyberinfrastructure: An Organization Adapting to Change \cite{10.1145/2145204.2145339}                                                   & 2012 & CSCW        \\ \hline
Repair Worlds: Maintenance, Repair, and ICT for Development in Rural Namibia \cite{10.1145/2145204.2145224}                                                            & 2012 & CHI         \\ \hline
Human infrastructure as process and effect: its impact on individual scientists’ participation in international collaboration \cite{10.1145/2389176.2389184}           & 2012 & GROUP       \\ \hline
Electronic medication management: a socio-technical change process in clinical practice \cite{10.1145/2145204.2145335}                                                 & 2012 & CSCW        \\ \hline
Infrastructural Experiences: An Empirical Study of an Online Arcade Game Platform in China \cite{10.1145/2145204.2145294}                                              & 2012 & CSCW        \\ \hline
Interacting with Infrastructure: A Case for Breaching Experiments in Home Computing Research \cite{10.1145/2145204.2145319}                                            & 2012 & CSCW        \\ \hline
Homework: Putting Interaction into the Infrastructure \cite{Mortier2012}                                                                                   & 2012 & UIST        \\ \hline
At Home with Agents: Exploring Attitudes towards Future Smart Energy Infrastructures \cite{10.1145/2470654.2466152}                                                    & 2013 & CHI         \\ \hline
The Collective Infrastructural Work of Electricity: Exploring Feedback in a Prepaid University Dorm in China \cite{10.1145/2493432.2493497}                            & 2013 & UbiComp     \\ \hline
Turkopticon: Interrupting worker invisibility in Amazon Mechanical Turk \cite{Irani2013}                                                                 & 2013 & CHI         \\ \hline
Infrastructure and Vocation: Field, Calling and Computation in Ecology \cite{10.1145/2470654.2481397}                                                                   & 2013 & CHI         \\ \hline
“Facebook is a Luxury”: An Exploratory Study of Social Media Use in Rural Kenya \cite{10.1145/2441776.2441783}                                                          & 2013 & CSCW        \\ \hline
“Everybody Knows What You’re Doing”: A Critical Design Approach to Personal Informatics \cite{10.1145/2470654.2466467}                                                 & 2013 & CHI         \\ \hline
Designing for Repair? Infrastructures and Materialities of Breakdown \cite{10.1145/2531602.2531692}                                                                    & 2014 & CSCW        \\ \hline
Learning, innovation, and sustainability among mobile phone repairers in Dhaka, Bangladesh \cite{Jackson2014}                                              & 2014 & DIS         \\\hline
The role of data in aligning the “unique identity” infrastructure in India \cite{10.1145/2531602.2531687}                                                              & 2014 & CSCW        \\\hline
The kernel of a research infrastructure \cite{10.1145/2531602.2531700}                                                                                                 & 2014 & CSCW        \\\hline
Reconciling Rhythms: Plans and Temporal Alignment in Collaborative Scientific Work \cite{10.1145/2531602.2531736}                                                      & 2014 & CSCW        \\\hline
Avocados Crossing Borders: The Missing Common Information Infrastructure for International Trade \cite{10.1145/2631488.2631500}                                         & 2014 & CABS        \\\hline
Towards Sociable Technologies: An Empirical Study on Designing Appropriation Infrastructures for 3D Printing \cite{10.1145/2598510.2598528}                            & 2014 & DIS         \\\hline
Ethnography of Scaling, or, How to a Fit a National Research Infrastructure in the Room \cite{10.1145/2531602.2531624}                                                 & 2014 & CSCW        \\\hline
A Story without End: Writing the Residual into Descriptive Infrastructure \cite{10.1145/2598510.2598553}                                                               & 2014 & DIS         \\\hline
Creating Sustainable Cyberinfrastructures \cite{10.1145/2702123.2702216}                                                                                               & 2015 & CHI         \\\hline
(Infra)Structures of Volunteering  \cite{10.1145/2675133.2675153}                                                                                                      & 2015 & CSCW        \\\hline
Standards and/as Innovation: Protocols, Creativity, and Interactive Systems Development in Ecology \cite{10.1145/2702123.2702564}                                      & 2015 & CHI         \\\hline
Why Replacing Legacy Systems Is So Hard in Global Software Development: An Information Infrastructure Perspective \cite{10.1145/2675133.2675232}                       & 2015 & CSCW        \\\hline
Anticipation Work: Cultivating Vision in Collective Practice \cite{10.1145/2675133.2675298}                                                                            & 2015 & CSCW        \\\hline
Restructuring Human Infrastructure: The Impact of EHR Deployment in a Volunteer-Dependent Clinic \cite{10.1145/2675133.2675277}                                         & 2015 & CSCW        \\\hline
Residual mobilities: Infrastructural displacement and post-colonial computing in Bangladesh \cite{Ahmed2015}                                             & 2015 & CHI         \\\hline
We Are Dynamo: Overcoming Stalling and Friction in Collective Action for Crowd Workers \cite{10.1145/2702123.2702508}                                                  & 2015 & CHI         \\\hline
Caring for Batteries: Maintaining Infrastructures and Mobile Social Contexts \cite{10.1145/2785830.2785864}                                                            & 2015 & MobileHCI   \\\hline
Hacking as transgressive infrastructuring: Mobile phonenetworks and the German chaos computer club \cite{Wagenknecht2016}                                      & 2016 & CSCW        \\\hline
Motivating Invisible Contributions: Framing Volunteer Classification Design in a Fanfiction Repository \cite{10.1145/2957276.2957295}                                  & 2016 & GROUP       \\\hline
Transition Resilience with ICTs \cite{10.1145/2858036.2858109}                                                                                                         & 2016 & CHI         \\\hline
Convivial Decay: Entangled Lifetimes in a Geriatric Infrastructure \cite{10.1145/2818048.2820077}                                                                      & 2016 & CSCW        \\\hline
Infrastructuring and the Challenge of Dynamic Seams in Mobile Knowledge Work \cite{10.1145/2818048.2820015}                                                            & 2016 & CSCW        \\\hline
Values in repair \cite{Houston2016}                                                                                                                        & 2016 & CHI         \\\hline
Logistics as Care and Control: An Investigation into the UNICEF Supply Division \cite{10.1145/2858036.2858503}                                                         & 2016 & CHI         \\\hline
Infrastructure in the wild: What mapping in post-earthquake Nepal reveals about infrastructural emergence \cite{10.1145/2858036.2858545}                               & 2016 & CHI         \\\hline
Breaking Down While Building Up: Design and Decline in Emerging Infrastructures \cite{10.1145/2858036.2858420}                                                         & 2016 & CHI         \\\hline
Happenstance, strategies and tactics: Intrinsic design in a volunteer-based community \cite{Bodker2016}                                                   & 2016 & NordiCHI    \\\hline
Open Data in Scientific Settings: From Policy to Practice \cite{10.1145/2858036.2858543}                                                                               & 2016 & CHI         \\\hline
Complex Decision-Making in Clinical Practice \cite{10.1145/2818048.2819952}                                                                                            & 2016 & CSCW        \\\hline
Technology literacy in poor infrastructure environments: Characterizing wayfinding strategies in Lebanon \cite{10.1145/2935334.2935352}                                & 2016 & MobileHCI   \\\hline
SOLE meets MOOC: Designing infrastructure for online self-organised learning with a social mission \cite{Celina2016}                                      & 2016 & DIS         \\\hline
On making data actionable: How activists use imperfect data to foster social change for human rights violations in Mexico \cite{Garcia2017}               & 2017 & CSCW        \\\hline
"Who Has Plots?": Contextualizing Scientific Software, Practice, and Visualizations \cite{10.1145/3134720}               & 2017 & CSCW        \\\hline
Crowdfunding Platforms and the Design of Paying Publics \cite{10.1145/3025453.3025979}                                                                                 & 2017 & CHI         \\\hline
Bots, Seeds and People: Web Archives as Infrastructure \cite{10.1145/2998181.2998345}                                                                                  & 2017 & CSCW        \\\hline
What lies above: Alternative user experiences produced through focusing attention on GNSS infrastructure \cite{10.1145/3064663.3064757}                                & 2017 & DIS         \\\hline
Growing the Blockchain Information Infrastructure \cite{10.1145/3025453.3025959}                                                                                       & 2017 & CHI         \\\hline
Infrastructure as Creative Action: Online Buying, Selling, and Delivery in Phnom Penh \cite{10.1145/3025453.3025889}                                                   & 2017 & CHI         \\\hline
Informality and Invisibility: Traditional Technologies as Tools for Collaboration in an Informal Market  \cite{10.1145/3025453.3025643}                                & 2017 & CHI         \\\hline
Notes on the concept of data interoperability: Cases from an ecology of AIDS research infrastructures \cite{10.1145/2998181.2998344}                                   & 2017 & CSCW        \\\hline
From margins to seams: Imbrication, inclusion, and torque in the Aadhaar identification project \cite{Singh2017}                                          & 2017 & CHI         \\\hline
Tap the “Make This Public” Button: A Design-Based Inquiry into Issue Advocacy and Digital Civics \cite{10.1145/3025453.3026034}                                         & 2017 & CHI         \\\hline
3D printers as sociable technologies: Taking appropriation infrastructures to the Internet of Things \cite{Ludwig2017}                                    & 2017 & TOCHI       \\\hline
The gig economy and information infrastructure: The case of the digital nomad community \cite{Sutherland2017}                                                 & 2017 & CSCW        \\\hline
Social media seamsters: Stitching platforms \& audiences into local crisis infrastructure \cite{10.1145/2998181.2998290}                                               & 2017 & CSCW        \\\hline
Enabling Polyvocality in Interactive Documentaries through “Structural Participation.” \cite{10.1145/3025453.3025606}                                                  & 2017 & CHI         \\\hline
Friction in Arenas of Repair: Hacking, Security Research, and Mobile Phone Infrastructure \cite{10.1145/2998181.2998308}                                                & 2017 & CSCW        \\\hline
Mapping Silences, Reconfiguring Loss \cite{10.1145/3274430} & 2018 & CSCW  \\\hline
Data Handling in Knowledge Infrastructures: A Case Study from Oil Exploration \cite{10.1145/3274392} & 2018 & CSCW  \\\hline
Designing for collaborative infrastructuring: Supporting resonance activities \cite{Ludwig2018} & 2018 & CSCW  \\\hline
How Latino Children in the U.S. Engage in Collaborative Online Information Problem Solving with Their Families \cite{10.1145/3274409} & 2018 & CSCW  \\\hline
Public WiFi is for Men and Mobile Internet is for Women: Interrogating Politics of Space and Gender around WiFi Hotspots \cite{10.1145/3274395} & 2018 & CSCW  \\\hline
Transforming Taxonomic Interfaces: "Arm?S Length" Cooperative Work and the Maintenance of a Long-Lived Classification System \cite{10.1145/3274442} & 2018 & CSCW  \\\hline
Infrastructural Inaccessibility: Tech Entrepreneurs in Occupied Palestine \cite{Bjorn2018} & 2018 & TOCHI  \\\hline
Infrastructural Grind: Introducing Blockchain Technology in the Shipping Domain \cite{10.1145/3148330.3148345}                                                          & 2018 & GROUP       \\\hline
Stitching Infrastructures to Facilitate Telemedicine for Low-Resource Environments \cite{10.1145/3173574.3173958}                                                      & 2018 & CHI         \\\hline
El Paquete semanal: The week’s internet in Havana \cite{10.1145/3173574.3174213}                                                                                       & 2018 & CHI         \\\hline
“More than Just Space”: Designing to Support Assemblage in Virtual Creative Hubs \cite{10.1145/3196709.3196758}                                                        & 2018 & DIS         \\\hline
Design Artefacts to Support People with a Disability to Build Personal Infrastructures \cite{10.1145/3196709.3196749}                                                  & 2018 & DIS         \\\hline
Crowdsourcing Rural Network Maintenance and Repair via Network Messaging \cite{10.1145/3173574.3173641}                                                                & 2018 & CHI         \\\hline
Infrastructuring the Solidarity Economy: Unpacking Strategies and Tactics in Designing Social Innovation \cite{10.1145/3173574.3174055}                                & 2018 & CHI         \\\hline
Fostering Commonfare. Infrastructuring Autonomous Social Collaboration \cite{10.1145/3173574.3174026}                                                                  & 2018 & CHI         \\\hline
Strategies for Engaging Communities in Creating Physical Civic Technologies \cite{10.1145/3173574.3174081}                                                             & 2018 & CHI         \\\hline
Trust and Technology Repair Infrastructures in the Remote Rural Philippines Navigating Urban-Rural Seams \cite{Jang2019}                                & 2019 & CSCW        \\\hline
“Mothers as Candy Wrappers”: Critical infrastructure supporting the transition into motherhood \cite{Britton2019}                                          & 2019 & GROUP       \\\hline
Making Healthcare Infrastructure Work: Unpacking the Infrastructuring Work of Individuals \cite{10.1145/3290605.3300688}                                               & 2019 & CHI         \\\hline
“Routine Infrastructuring” as “Building Everyday Resilience with Technology”: When Disruption Becomes Ordinary \cite{10.1145/3359175}                          & 2019 & CSCW        \\\hline
Designing for the infrastructure of the supply chain of Malay handwoven songket in Terengganu \cite{10.1145/3290605.3300716}                                           & 2019 & CHI         \\\hline
Guerilla Warfare and the Use of New (and Some Old) Technology: Lessons from FARC’s Armed Struggle in Colombia \cite{10.1145/3290605.3300810}                           & 2019 & CHI         \\\hline
Blockchain assemblages whiteboxing technology and transforming infrastructural imaginaries  \cite{10.1145/3290605.3300496}                                             & 2019 & CHI         \\\hline
Infrastructuring the imaginary how sea-level rise comes to matter in the San Francisco Bay area \cite{Soden2019}                                         & 2019 & CHI         \\\hline
Moving into a technology land: Exploring the challenges for the Refugees in Canada in Accessing its Computerized Infrastructures \cite{Sabie2019}        & 2019 & COMPASS     \\\hline
“Parar-Daktar Understands My Problems Better”: Disentangling the Challenges to Designing Better Access to Healthcare in Rural Bangladesh \cite{10.1145/3359270} & 2019 & CSCW        \\\hline
Infrastructuring public service transformation: Creating collaborative spaces between communities and institutions through HCI research \cite{10.1145/3310284} & 2019 & TOCHI       \\\hline
Infrastructuring Food Democracy: The Formation of a Local Food Hub in the Context of Socio-Economic Deprivation  \cite{10.1145/3359159}                        & 2019 & CSCW        \\\hline
Precarious interventions: Designing for ecologies of care \cite{Kaziunas2019} & 2019 & CSCW        \\\hline
The Coerciveness of the Primary Key: Infrastructure Problems in Human Services Work \cite{10.1145/3359153} & 2019 & CSCW        \\\hline
HOPE for Computing Education: Towards the Infrastructuring of Support for University-School Partnerships  \cite{10.1145/3290605.3300729}                               & 2019 & CHI         \\\hline
Workshops as Boundary Objects for Data Infrastructure Literacy and Design \cite{10.1145/3322276.3322330}                                                               & 2019 & DIS         \\\hline
Designing with Waste: A Situated Inquiry into the Material Excess of Making \cite{10.1145/3322276.3322320}                                                             & 2019 & DIS         \\\hline
The social infrastructure of Co-spaces: Home, work, and sociable places for digital nomads  \cite{Lee2019}                                             & 2019 & CSCW        \\\hline
BlocKit: A Physical Kit for Materializing and Designing for Blockchain Infrastructure \cite{10.1145/3322276.3322370}                                                   & 2019 & DIS         \\\hline
“Like Shock Absorbers”: Understanding the Human Infrastructures of Technology-Mediated Mental Health Support \cite{Pendse2020}                           & 2020 & CHI         \\\hline
User’s Role in Platform Infrastructuralization: WeChat as an Exemplar \cite{Zhou2020}                                                                   & 2020 & CHI         \\\hline
Gaming the Name: Player Strategies for Adapting to Name Constraints in Online Videogames \cite{10.1145/3410404.3414259}                                                & 2020 & CHI PLAY    \\\hline
The Social Network: How People with Visual Impairment Use Mobile Phones in Kibera, Kenya \cite{10.1145/3313831.3376658}                                                 & 2020 & CHI         \\\hline
“I Can’t even Buy Apples if i Don’t Use Mobile Pay?”: When Mobile Payments Become Infrastructural in China \cite{Shen2020}                              & 2020 & CSCW        \\\hline
'Yes, I Comply!': Motivations and Practices around Research Data Management and Reuse across Scientific Fields \cite{10.1145/3415212} & 2020 & CSCW        \\\hline
Infrastructural Speculations: Tactics for Designing and Interrogating Lifeworlds \cite{10.1145/3313831.3376515}                                                        & 2020 & CHI         \\\hline
An Internet-Less World? Expected Impacts of a Complete Internet Outage with Implications for Preparedness and Design \cite{10.1145/3375183}                    & 2020 & GROUP       \\\hline
Examining Opaque Infrastructures with the Desktop Odometer \cite{Viny2021}                                                                              & 2021 & DIS         \\\hline
Breakdowns and Breakthroughs: Observing Musicians’ Responses to the COVID-19 Pandemic \cite{Cai2021}                                                   & 2021 & CHI         \\\hline
Cracking Public Space Open \cite{Kozubaev2021}                                                                                                              & 2021 & CHI         \\\hline
Medical maker response to covid-19: Distributed manufacturing infrastructure for stopgap protective equipment \cite{Lakshmi2021}                           & 2021 & CHI         \\\hline
Contact Zones: Designing for More-than-Human Food Relations \cite{10.1145/3449121}  & 2021 & CSCW        \\\hline
The Labor of Maintaining and Scaling Free and Open-Source Software Projects \cite{10.1145/3449249} & 2021 & CSCW        \\\hline
Seeing Like an Infrastructure: Low-Resolution Citizens and the Aadhaar Identification Project \cite{10.1145/3476056} & 2021 & CSCW        \\\hline
"They Can Only Ever Guide": How an Open Source Software Community Uses Roadmaps to Coordinate Effort \cite{10.1145/3449232} & 2021 & CSCW        \\\hline
Data Integration as Coordination: The Articulation of Data Work in an Ocean Science Collaboration \cite{10.1145/3432955} & 2021 & CSCW        \\\hline
What’ s in a Network ? Infrastructures of Mutual Aid for Digital Platform Workers during COVID-19 \cite{Qadri2021} & 2021 & CSCW        \\\hline
Parsing the ‘Me’ in \# MeToo: Sexual Harassment, Social Media, and Justice Infrastructures \cite{Moitra2021}                                              & 2021 & CSCW        \\\hline
“As a Squash Plant Grows”: Social Textures of Sparse \cite{Duarte2021}                                                                                 & 2021 & TOCHI       \\\hline
Biographies of biometric devices: The POS machine at work in India’s PDS \cite{Mortier2012}                                                                & 2021 & CHI         \\\hline
Un Grano de Arena: Infrastructural Care, Social Media Platforms, and the Venezuelan Humanitarian Crisis \cite{10.1145/3432946}                                 & 2021 & CSCW        \\\hline
Hugs, Bible Study, and Speakeasies: Designing for Older Adults’ Multimodal Connectedness \cite{Richards2021}                                            & 2021 & DIS         \\\hline
CrowdSolve: Managing Tensions in an Expert-Led Crowdsourced Investigation \cite{Venkatagiri2021}                                                              & 2021 & CSCW        \\\hline
Cat and Mouse Game: Patching Bureaucratic Work Relations by Patching Technologies \cite{Veeraraghavan2021}                                                      & 2021 & CSCW        \\\hline
Teachers’ perceptions around digital games for children in low-resource schools for the blind \cite{India2021}                                           & 2021 & CHI         \\\hline
Leaving the field: Designing a socio-material toolkit for teachers to continue to design technology with children \cite{Scheepmaker2021}                       & 2021 & CHI \\\hline
“There Should Be More Than One Voice in a Healthy Society” : Infrastructural Violence and Totalitarian Computing in China \cite{Li2021} & 2021 & CSCW \\\hline
Infrastructuring Telehealth in (In) Formal Patient-Doctor Contexts \cite{Bhat2021} & 2021 & CSCW \\\hline
The Flaky Accretions of Infrastructure: Sociotechnical Systems, Citizenship, and the Water Supply \cite{Joshi2021} & 2021 & CSCW \\\hline
The Pandemic Shift to Remote Learning under Resource Constraints \cite{Ravi2021} & 2021 & CSCW \\\hline
The Work of Infrastructural Bricoleurs in Building Civic Data Dashboards \cite{10.1145/3512971} & 2022 & CSCW \\\hline
Disinformation as Infrastructure: Making and Maintaining the QAnon Conspiracy on Italian Digital Media \cite{10.1145/3512931} & 2022 & CSCW \\\hline
Making Space for Cultural Infrastructure : The Breakdown and Maintenance Work of Independent Movie Theaters During Crisis \cite{10.1145/3491102.3501840} & 2022 & CHI \\\hline
“Hartal (Strike) Happens Here Everyday”: Understanding Impact of Disruption on Education in Kashmir \cite{10.1145/3491102.3502126} & 2022 & CHI \\\hline
The Village: Infrastructuring Community-Based Mentoring to Support Adults Experiencing Poverty \cite{10.1145/3491102.3501949} & 2022 & CHI \\\hline
Seamless Visions, Seamful Realities: Anticipating Rural Infrastructural Fragility in Early Design of Digital Agriculture \cite{10.1145/3491102.3517579}  & 2022 & CHI \\\hline
Cultural Influences on Chinese Citizens’ Adoption of Digital Contact Tracing: A Human Infrastructure Perspective \cite{10.1145/3491102.3517572} & 2022 & CHI \\\hline
Putting the Waz on Social Media: Infrastructuring Online Islamic Counterpublic through Digital Sermons in Bangladesh \cite{10.1145/3491102.3502006} & 2022 & CHI \\\hline
Shifting Trust: Examining How Trust and Distrust Emerge, Transform, and Collapse in COVID-19 Information Seeking \cite{10.1145/3491102.3501889} & 2022 & CHI \\\hline
Care Infrastructures for Digital Security in Intimate Partner Violence \cite{10.1145/3491102.3502038} & 2022 & CHI \\\hline
Designing Flexible Longitudinal Regimens: Supporting Clinician Planning for Discontinuation of Psychiatric Drugs \cite{10.1145/3491102.3502206} & 2022 & CHI \\\hline
Blockchain and Beyond: Understanding Blockchains through Prototypes and Public Engagement \cite{Murray-Rust2022} & 2022 & TOCHI \\\hline
"What is your envisioned future?": Toward human-AI enrichment in data work of asthma care \cite{10.1145/3555157} & 2022 & CSCW \\\hline
Counting to be counted: Anganwadi workers and digital infrastructures of ambivalent care \cite{spCountingBeCounted2022} & 2022 & CSCW \\\hline
There is no app for that: Manifestations of the digital divides during COVID-19 school closures in India \cite{dhaygudeThereNoApp2022} & 2022 & CSCW \\\hline
The Chinese diaspora and the attempted WeChat ban: Platform precarity, anticipated impacts, and infrastructural migration \cite{10.1145/3555122} & 2022 & CSCW \\\hline

Speculative vulnerability: Uncovering the temporalities of vulnerability in people's experiences of the pandemic \cite{10.1145/3555586} & 2022 & CSCW \\\hline

Human and technological infrastructures of fact-checking \cite{10.1145/3555143} & 2022 & CSCW \\\hline

Nationalizing the Internet to break a protest movement: Internet shutdown and counter-appropriation in Iran of late 2019 \cite{grinkoNationalizingInternetBreak2022} & 2022 & CSCW \\\hline

Revolting from abroad: The formation of a lebanese transnational public \cite{10.1145/3555131} & 2022 & CSCW \\\hline

Gig platforms as faux infrastructure: a case study of women beauty workers in India \cite{anjalianwarGigPlatformsFaux2022} & 2022 & CSCW \\\hline

"We dream of climbing the ladder; to get there, we have to do our job better": Designing for Teacher Aspirations in rural Cote d'Ivoire \cite{10.1145/3530190.3534794} & 2022 & COMPASS \\\hline

Blockchain and beyond: Understanding blockchains through prototypes and public engagement \cite{10.1145/3503462} & 2023 & TOCHI \\\hline

Infrastructures for virtual volunteering at online music festivals \cite{10.1145/3579498} & 2023 & CSCW \\\hline

Lessons learned from a comparative study of long-term action research with community design of infrastructural systems \cite{10.1145/3579502} & 2023 & CSCW \\\hline

Organizing oceanographic infrastructure: The work of making a software pipeline repurposable \cite{10.1145/3579512} & 2023 & CSCW \\\hline

"Hey, can you add captions?": The critical infrastructuring practices of neurodiverse people on TikTok \cite{10.1145/3579490} & 2023 & CSCW \\\hline

"We picked community over privacy": Privacy and Security Concerns Emerging from Remote Learning Sociotechnical Infrastructure During COVID-19 \cite{wagmanWePickedCommunity2023} & 2023 & CSCW \\\hline

Participatory noticing through photovoice: Engaging arts- and community-based approaches in design research \cite{10.1145/3563657.3596041} & 2023 & DIS \\\hline

Understanding Human Intervention in the Platform Economy: A Case Study of an Indie Food Delivery Service \cite{10.1145/3544548.3581517} & 2023 & CHI \\\hline

Infrastructural work behind the scene: a study of formalized peer-support practices for mental health \cite{10.1145/3544548.3580657} & 2023 & CHI \\\hline

Shifting from surveillance-as-safety to safety-through-noticing: a photovoice study with eastside detroit residents \cite{10.1145/3544548.3581474} & 2023 & CHI \\\hline

Infrastructuring care: How trans and non-binary people meet health and well-being needs through technology \cite{10.1145/3544548.3581040} & 2023 & CHI \\\hline

Commoning as a strategy for HCI research and design in south asia \cite{10.1145/3613904.3642547} & 2024 & CHI \\\hline

ml-machine.org: Infrastructuring a research product to disseminate AI literacy in education \cite{10.1145/3613904.3642539} & 2024 & CHI \\\hline

Not just a dot on the map: Food delivery workers as infrastructure \cite{10.1145/3613904.3641918} & 2024 & CHI \\\hline

"Obviously, nothing's gonna happen in five minutes": How adolescents and young adults infrastructure resources to learn type 1 diabetes management \cite{10.1145/3613904.3642612} & 2024 & CHI \\\hline

"Vulnerable, Victimized, and Objectified": Understanding Ableist Hate and Harassment Experienced by Disabled Content Creators on Social Media \cite{heung_vulnerable_2024} & 2024 & CHI \\\hline

Seam work and simulacra of societal impact in networking research: a critical technical practice approach \cite{10.1145/3613904.3642337} & 2024 & CHI \\\hline

Aftermath: Infrastructure, resources, and organizational adaptation in the wake of disaster \cite{10.1145/3637294} & 2024 & CSCW \\\hline

"Because Some Sighted People, They Don't Know What the Heck You're Talking About:" A Study of Blind TikTokers' Infrastructuring Work to Build Independence \cite{lyu_because_2024} & 2024 & CSCW \\\hline

Concept of operations as epistemic object: The sociotechnical design roles of a systems engineering document \cite{10.1145/3637311} & 2024 & CSCW \\\hline

Entangled amid misaligned seams: Limitations to technology-mediated care for repairing infrastructural breakdowns in a youth empowerment program \cite{10.1145/3686896} & 2024 & CSCW \\\hline

Infrastructuring community fridges for food commoning \cite{10.1145/3637352} & 2024 & CSCW \\\hline

Navigating the job-seeking journey: Challenges and opportunities for digital employment support in kashmir \cite{10.1145/3637375} & 2024 & CSCW \\\hline

Reconfiguring data relations: Institutional dynamics around data in local governance \cite{10.1145/3686959} & 2024 & CSCW \\\hline

Security patchworking in lebanon: Infrastructuring across failing infrastructures \cite{10.1145/3637397} & 2024 & CSCW \\\hline

Socio-digital rural resilience: An exploration of information infrastructures within and across rural villages during covid-19 \cite{10.1145/3637400} & 2024 & CSCW \\\hline

Towards inclusive futures for worker wellbeing \cite{10.1145/3637414} & 2024 & CSCW \\\hline

\end{longtable}
\end{center}

\end{document}